\documentclass[twocolumn,prb,superscriptaddress]{revtex4-1}

\usepackage{graphicx}
\usepackage{dcolumn}
\usepackage{bm}
\usepackage{hyperref}
\usepackage{amsmath}
\usepackage{amssymb}
\usepackage[normalem]{ulem}
\usepackage[usenames,dvipsnames,svgnames,table]{xcolor}

\begin{document}

\preprint{APS/123-QED}

\title{Interplay between quantum confinement and dielectric mismatch\\ for ultra-shallow dopants.}

\author{J.A. Mol}
\affiliation{Centre for Quantum Computation and Communication Technology, University of New South Wales, Sydney NSW 2052, Australia}
\affiliation{Kavli Institute of Nanoscience, Delft University of Technology,
Lorentzweg 1, 2628 CJ Delft, The Netherlands}
\author{J. Salfi}
\affiliation{Centre for Quantum Computation and Communication Technology, University of New South Wales, Sydney NSW 2052, Australia}
\author{J.A. Miwa}
\affiliation{Centre for Quantum Computation and Communication Technology, University of New South Wales, Sydney NSW 2052, Australia}
\author{M.Y. Simmons}
\affiliation{Centre for Quantum Computation and Communication Technology, University of New South Wales, Sydney NSW 2052, Australia}
\author{S. Rogge}
\affiliation{Centre for Quantum Computation and Communication Technology, University of New South Wales, Sydney NSW 2052, Australia}
\affiliation{Kavli Institute of Nanoscience, Delft University of Technology,
Lorentzweg 1, 2628 CJ Delft, The Netherlands}

\date{\today}
             
\begin{abstract}
Understanding the electronic properties of dopants near an interface is a critical challenge for nano-scale devices. We have determined the effect of dielectric mismatch and quantum confinement on the ionization energy of individual acceptors beneath a hydrogen passivated silicon (100) surface. Whilst dielectric mismatch between the vacuum and the silicon at the interface results in an image charge which enhances the binding energy of sub-surface acceptors, quantum confinement is shown to reduce the binding energy. Using scanning tunneling spectroscopy we measure resonant transport through the localized states of individual acceptors. Thermal broadening of the conductance peaks provides a direct measure for the absolute energy scale. Our data unambiguously demonstrates that these two independent effects compete with the result that the ionization energy is less than 5~meV lower than the bulk value for acceptors less than a Bohr radius from the interface.\end{abstract}

\pacs{Valid PACS appear here}

\maketitle

\section{Introduction}
The operation of semiconductor devices is based on the possibility to locally change the electron properties of the host material by means of doping. As device dimensions continue to decrease, the surface-to-volume ratio of active channels increases and the effect of the semiconductor-insulator interface on local doping starts to dominate device properties \cite{Bukhori:2010ka}. Previous studies have suggested that dielectric mismatch at the semiconductor-insulator interface leads to an increase in ionization energy of dopants near the interface \cite{Bjork:2009bi,Pierre:2009db}. In silicon nanowires this leads to doping deactivation and consequently an increase of resistivity with decreasing diameter\cite{Diarra:2007ij,Bjork:2009bi}. However, recent transport spectroscopy experiments on single arsenic donors in gated nanowires did not report an appreciable increase in ionization energy \cite{Sellier:2006wt}. These results appear contradictory.

The ionization energies of shallow donor and acceptor impurities are qualitatively described by effective mass theory \cite{Luttinger:1955ee}, which works especially well for light impurity atoms such as Li and B. Since the Coulomb potential is strongly screened due to the polarization of the semiconductor, the ionization energy of dopant impurities is only in the order of tens of meV. This simple picture breaks down in the presence of an interface. Dielectric mismatch between the semiconductor material and its surroundings is predicted to enhance the ionization energy \cite{Diarra:2007ij}. Moreover, for nanowires it is well known that when the thickness of the nanowire approaches the Bohr radius of the impurity the ionization energy \emph{increases} due to quantum confinement \cite{Loudon:1959ih,Bryant:1984iy}. However, in the case of a half-space, i.e. a flat interface, effective mass theory predicts a \emph{decrease} in the ionization energy due to quantum confinement \cite{Hao:2009eq,Calderon:2010kz}. As a result of these two competing effects, dielectric mismatch and quantum confinement, the ionization energy of dopant near a flat interface is expected to be bulk-like \cite{Hao:2009eq,Calderon:2010kz}.

Here, we use low-temperature (4.2 K) scanning tunneling spectroscopy (STS) to directly measure the ionization energy of boron acceptors beneath the hydrogen terminated Si(100) surface ($N_A \sim 8\times10^{18}$~cm$^{-3}$). Experiments were performed with an ultra-high vacuum STM at liquid helium temperature $T=4.2$~K, this temperature is measured at the sample stage. A hydrogenated Si(100):H surface was prepared by flash annealing the sample to 1200~$^\circ$C three times for an integrated anneal time of 30 s followed by slow cool down from 850~$^\circ$C to 350~$^\circ$C. The sample was then exposed to 6 Langmuir of atomic hydrogen in order to hydrogen-passivate the surface. Previous studies on GaAs(110) \cite{Yakunin:2007dv,Loth:2008tc,Garleff:2010fp,Wijnheijmer:2010uc,Lee:2010ko}, InAs(110) \cite{Marczinowski:2007dt} and ZnO(0001) \cite{Zheng:2012ef} surfaces have proven that STS is a powerful tool to study sub-surface impurities. Using scanning tunneling microscopy the surroundings of each individual dopant atom can be imaged  and therefore any effect of dopant clustering or interface disorder, such as charge traps, on the ionization energy of the acceptors may be excluded. By analyzing the line shape of differential conductance within well known single-electron spectroscopy formalisms, we conclude that transport is predominantly thermally broadened. Consequently, the hole reservoir temperature is used as a reference to calibrate the coupling between the applied bias voltage and the potential landscape at the semiconductor-vacuum interface. In such a way we are able to obtain a direct measure for the acceptor ionization energy. Moreover, STS allows us to determine the distance of individual acceptors to the interface by measuring the spectral shift of the valence band due to the negatively charged acceptor nucleus. Importantly, from the thermally broadened single-electron transport through the localized acceptor state, in conjunction with the spectral shift of the valence band due to the ionized nucleus, the dopant depth can be directly correlated to its ionization energy.

All parameters for the determination of the depth and ionization energy of individual sub-surface acceptors are experimentally obtained from STS measurements. The voltage of ionization is directly determined from the onset of resonant tunnelling through the localized acceptor state, i.e. the ionization voltage coincides with the center of the differential conductance peak corresponding to this tunnel process. The tip induced band bending is inferred from the apparent shift of the onset of the valence and conduction band with respect to the known onsets in absence of the tip. Using the model of voltage dependent band bending introduced by Feenstra the tip voltage for flat-band conditions at the surface can be deduced. The lever arm between the applied bias voltage and the energy shift of the localized acceptor levels is determined from fitting the differential conductance peaks, which result from resonant tunnelling through the acceptor states, to a thermally broadened Lorentzian line shape that is stretched by the lever arm. Finally, the depth of each acceptor is obtained from a fit of the lateral, spatial dependence of the valence band onset around the dopant where the depth and the effective dielectric constant are used as two, independent, fitting parameters. For five individual acceptors up to 2 nm below the surface, we find that the ionization energy decreases less than 5 meV from the known bulk value of 45 meV.\\

\begin{figure}
\includegraphics[width=90mm]{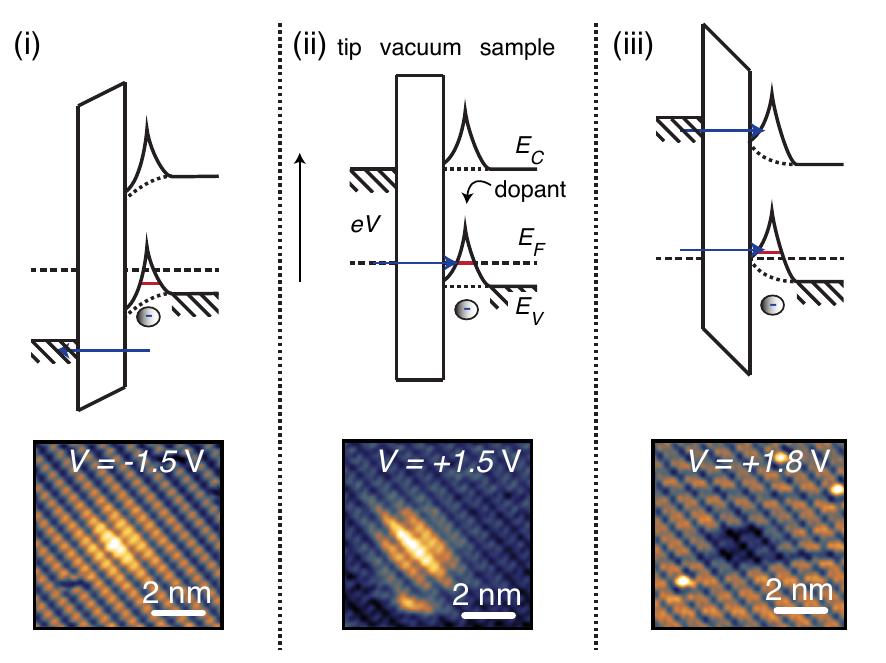}
\caption{\label{prx_fig1} Schematic energy diagram of the tip-vacuum-sample tunnel junction. (i) When $eV<E_V<eV_{FB}$ electrons tunnel from the valence band to the tip and the presence of a sub-surface acceptor is observed as a protrusion in the STM topography. (ii) In case $eV_{FB}<eV\sim E_C$ direct tunneling via the localized acceptor state leads to a protrusion in the topography. (iii) Suppression of the local density of states by the acceptor potential reduces the tunnel current when $E_C<eV$, resulting in a depression in the topography.}
\end{figure}

\section{Results and discussion}
The schematic energy diagrams in Fig.~\ref{prx_fig1} illustrate three different transport regimes in which we study the sub-surface boron acceptors: (i) charge sensing in the valence band, (ii) resonant tunneling through the localized acceptor state and (iii) charge sensing in the conduction band. When a negative sample bias voltage is applied, the presence of a sub-surface acceptor results in an increase in the direct tunneling into the valence band due to an increase in the local density of states (LDOS) caused by the negatively charged nucleus of the acceptor. The increase in tunnel current leads to a height increase in the STM topography, as is shown in figure~\ref{prx_fig1}(i). Likewise, when a positive voltage is applied, suppression of the local density of states by the acceptor potential leads to a decrease in the direct tunnel current into the conduction band and resulting in a dip in the STM topography \cite{Marczinowski:2007dt}(Fig.~\ref{prx_fig1}(iii)). However, when a positive voltage is applied such that the Fermi energy of the tip is close to the conduction band edge, the transport is no longer dominated by direct tunneling into the conduction band but by resonant transport through the localized acceptor state (Fig~\ref{prx_fig1}(ii)). 

Since the Fermi energy of the heavily B-doped sample is pinned at the bulk acceptor level $E_{A}^{bulk}$, the voltage at which the localized acceptor state is equal to the Fermi level, $V_{onset}$, with respect to the flat-band voltage, $V_{FB}$, is a direct measure for the difference in ionization energy, $\Delta E=E_A-E_A^{bulk}$, of the sub-surface acceptor with respect to the bulk ionization energy. The bias dependence of the energy level of the localized acceptor state $E_A$ is described by the lever arm $\alpha=e^{-1}dE_A/dV$  \cite{Wijnheijmer:2009il,Wijnheijmer:2011bg}. Here we present a direct measurement of the lever arm and measure the shift $\Delta E = -\alpha e (V_{onset}-V_{FB})$ by studying transport through individual sub-surface acceptors where we directly determine: (i) the potential due to the negatively charged nucleus, (ii) the flat-band voltage and depth of the acceptors from direct tunneling to/from the conductance/valence band and (iii) the ionization voltage and lever arm from single-electron transport through the localized acceptor state.\\

\begin{figure}
\includegraphics[width=90mm]{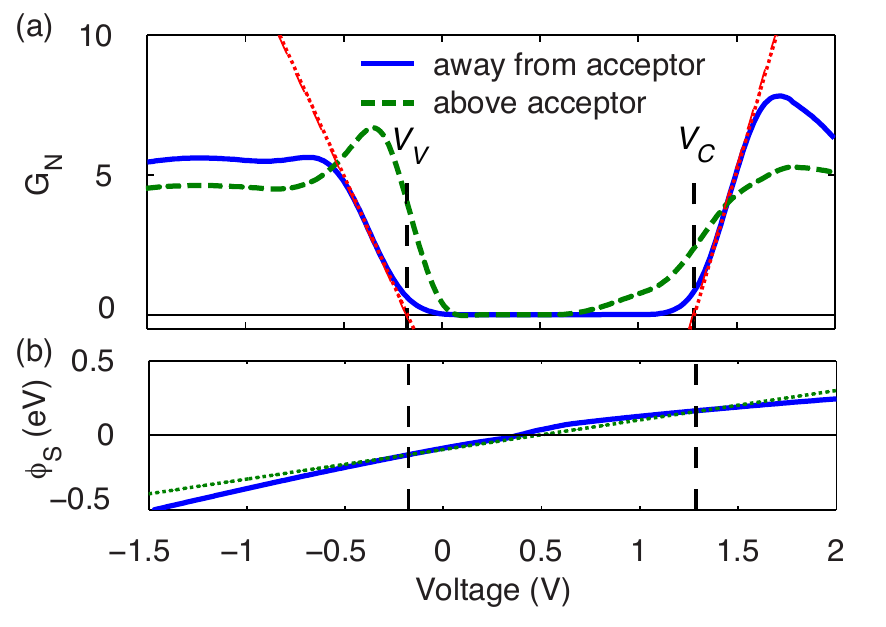}
\caption{\label{prx_fig2}(a) Normalized conductance $G_N$ measured on a hydrogen terminated Si(100):H surface away from (solid line) and above (dashed line) an acceptor. The top of the valence band $V_V$ and bottom of the conduction band $V_C$ are determined by the slopes of $G_N$ away from the acceptor (dotted lines). (b) Surface potential $\phi_S$ assuming linear TIBB (dotted line) and non-linear TIBB (solid line) calculated using the code developed by Feenstra \cite{Feenstra:2003ua}. The relevant parameters chosen for these calculations to match the change in band gap are the tip-sample separation 0.8~nm, the tip radius 8~nm and the doping concentration $1\times10^{18}$~cm$^{-3}$. Note that the doping concentration is lower than the substrate doping, which is expected from acceptor out-diffusion during the flash anneal.}
\end{figure}

Figure~\ref{prx_fig2}(a) shows the normalised conductance $G_{N} = (dI/dV)/(\overline{I/V})$  measured away from any sub-surface acceptor ($dI/dV$ is deduced numerically). The flat-band voltage $V_{FB}$ was extracted by comparing the 4.2~K bandgap of Si(001):H with voltages $V_V$ and $V_C$ for tunneling into the valence and conductance band edges (dashed lines in Fig.~\ref{prx_fig2}(a) and (b)). A first approximation of $V_{FB}$ is made by taking the tip induced band bending (TIBB) to be linear (dotted line in Fig.~\ref{prx_fig2}(b)). This constraint is subsequently relaxed in order to account for screening (solid line in Fig.~\ref{prx_fig2}(b)). Finally, the flat-band condition was independently measured from the apparent barrier height. 

The onset voltage for tunneling from the valence band $V_V$ and tunneling into the conduction band $V_C$ is determined by finding the voltage axis intercept (i.e., $G_{N}=0$) of the linear extrapolation of the $G_N-V$ curve at its maximum slope point (dotted lines in Fig.~\ref{prx_fig2}(a)) \cite{Feenstra:1994ur}. The potential $\phi_S(V)$ at the interface as a function of the applied sample voltages is obtained by from the flat-band energies $E_F-E_V=E_A^{bulk}-E_V=0.045$~eV \cite{Ramdas:1981tr} and $E_g=E_C-E_V=1.17$~eV at $T=4.2$~K \cite{Kittel:2004vfa} away from the acceptor. Assuming a linear relationship between the applied voltage and the potential at the interface
\begin{equation}
\frac{d\phi_S}{dV} = \frac{e(V_C-V_V)-E_g}{V_C-V_V},
\end{equation}
yields an approximated flat-band voltage $V_{FB}=0.5\pm0.1$~V which is inferred from the condition for tunneling from the valence band $\phi_S(V_V) = eV_V$
\begin{equation}
\frac{d\phi_S}{dV}(V_V-V_{FB})=eV_V.
\end{equation}
When the flat-band voltage lies within the band gap the $d\phi_S/dV$ can not be assumed linear as it is well known that for $V>V_{FB}$ accumulated carriers at the surface will screen the electric field from the tip more strongly. In order to correct for this effect we use the $\phi_S(V)$ calculated using the method of Feenstra \cite{Feenstra:2003ua} (Fig.~\ref{prx_fig2}(b)) and determine a corrected flat-band voltage $V_{FB}=0.38\pm0.1$~V. This flat-band voltage is smaller than expected from the difference between the bulk workfunction $\Phi=4.55$~eV \cite{Hopkins:2002ib} of tungsten and sample electron affinity $\chi=4.05$~eV \cite{Sze:2006uh}. The measured flat-band voltage corresponds to a tip workfunction $\Phi_{tip}=4.9\pm0.1$~eV (larger values for the tip workfunction have been previously reported). Finally, we independently confirmed the value of the tip workfunction by measuring the apparent barrier height \cite{Loth:2007us,Wijnheijmer:2010uc} resulting in $\Phi_{tip} = 4.8 \pm 0.1$~eV (see Appendix A).

\begin{figure}
\includegraphics[width=90mm]{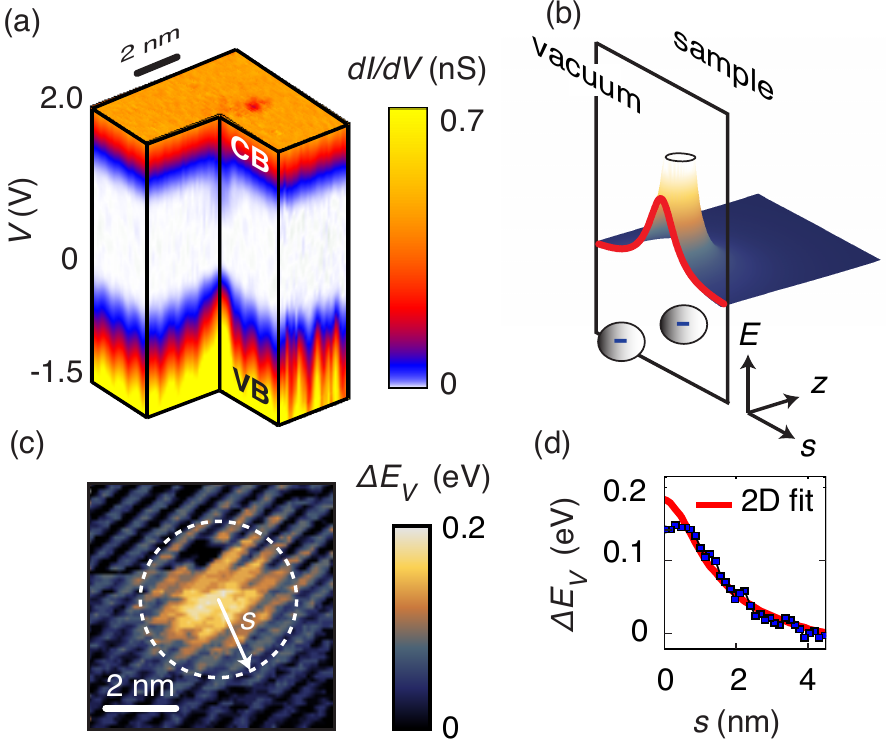}
\caption{\label{prx_fig3} (a) $dI/dV$ map recorded simultaneously with empty-state topography (+2.2~V, 300~pA). Tunneling from the valence band to the tip is indicated by VB, tunneling from the tip to the conduction band is indicated by CB. The $dI/dV$ map is cut at the acceptor site to show the upward shift of the valence band states. (b) Schematic potential landscape due to a negatively charged nucleus below the sample surface and its image charge in the vacuum. The solid line indicates the potential along the vacuum-semiconductor interface. (c) Shift in the valence band maximum $E_V$ as function of lateral tip position. (d) Azimuthal averaged shift in the valence band maximum $E_V$ (filled squares) as function lateral tip separation $s$ from the acceptor as indicated in (c). The dopant depth is determined by fitting $E_V$ to a bare Coulomb potential (solid line), taking into account the image charge due to dielectric mismatch as illustrated in (b).}
\end{figure}
The distance of the sub-surface acceptors to the interface is measured from the spectral shift of the valence band edge. Figure~\ref{prx_fig3}(a) shows the differential conductance ($dI/dV$) map, as a function of position and sample bias voltage $V$, measured simultaneously with the empty state topography at $V=2$~V shown in Fig.~\ref{prx_fig3}(a). At the acceptor site the $dI/dV$ map clearly shows an upward spectral shift of the valence band states due to the buried acceptor. The spatially resolved shift of the valence-band edge $\Delta E_V$ as shown in Fig.~\ref{prx_fig3}(b) is defined from the slope of $G_N$ as before. The first-order perturbation to the binding energy of the valence band states at the interface, $\psi_s$, and thus the shift of the valence-band edge, $\Delta E_V$, due to the potential of the negatively charged acceptor nucleus at position $r_0$ can be estimated as
\begin{equation}
\Delta E_V = \langle\psi_s |U_A(\vec{r},\vec{r}_0) |\psi_s\rangle,
\end{equation}
where $U_A(\vec{r},\vec{r}_0)$ is the acceptor potential at position $\vec{r}$. For an acceptor in bulk the impurity potential is given by \cite{Kobayashi:1996iu}
\begin{equation}
\label{eq:UA}
U^{bulk}_A(\vec{r},\vec{r}_0) = \frac{e_0^2}{4\pi\epsilon_0\epsilon_{\text{Si}}}\frac{e^{-k_0r}}{|\vec{r}-\vec{r}_0|},
\end{equation}
where $e_0$ is the electron charge and $k_0^{-1}$ the free-carrier screening length. When the semiconductor is depleted of free charge carriers, which is the case for $V<V_{FB}$, $k_0\rightarrow0$ and Eq.~\ref{eq:UA} will approach the dielectric-screened Coulomb potential \cite{Schechter:1962kw}. A finite $k_0$ will result in a shallower acceptor potential.

The presence of an interface will result in a change in dielectric screening of the impurity potential. An analytic solution to the Poisson equation at an interface can be found using the well-known method of image charges. For a planar tip-vacuum-silicon interface the potential in the semiconductor due to the ionised acceptor nucleus is given by \cite{Kumagai:1989ej,Hao:2009eq}: 
\begin{equation}
\label{eq:3layer}
\begin{split}
U_A(\vec{r},\vec{r}_0) = & \frac{e^2_0}{4\pi\epsilon_0\epsilon_{\text{Si}}}\Bigg[ \frac{1}{|\vec{r}-\vec{r}_0|} -\frac{\epsilon_{\text{v}}-\epsilon_{\text{Si}}}{\epsilon_{\text{v}}+\epsilon_{\text{Si}}}\frac{1}{|\vec{r}-\vec{r}_1|} \\
& + \sum_{n=0}^{\infty}\frac{4\epsilon_{\text{v}}\epsilon_{\text{S}i}}{(\epsilon_{\text{v}}+\epsilon_{\text{Si}})^2}\frac{\epsilon_{\text{v}}-\epsilon_{\text{tip}}}{\epsilon_{\text{v}}+\epsilon_{\text{tip}}}\xi^n\frac{1}{|\vec{r}-\vec{r}_{-2n-1}|}\Bigg],
\end{split}
\end{equation}
where 
\begin{equation}
\xi=\frac{\epsilon_{\text{v}}-\epsilon_{\text{Si}}}{\epsilon_{\text{v}}+\epsilon_{\text{Si}}}\frac{\epsilon_{\text{v}}-\epsilon_{\text{tip}}}{\epsilon_{\text{v}}+\epsilon_{\text{tip}}},
\end{equation}
and the charges are located at a distance
\begin{equation}
\begin{split}
z_0 = d,  \qquad &z_1 = -d, \\
z_{-2n-1} = -(2n+1)l-d, \quad &n=0,1,2,\dots
\end{split}
\end{equation}
from the vacuum-silicon interface, here $d$ is the distance of the acceptor nucleus to the interface and $l$ the tip-sample separation. It is important to note that when $l\gg d$, or indeed when $\epsilon_{tip}=\epsilon_v$, only the first two terms of Eq.~\ref{eq:3layer} remain, that is the ionized nucleus at $z=d$ and a single image charge at $z=-d$. As for free-carrier screening in the substrate, a finite tip-sample separation and $\epsilon_{tip}>\epsilon_{v}$ the presence of the tip will lead to a shallower acceptor potential.

In the case of a classical half-space, i.e. in the absence of screening by free carriers in the tip or the substrate, the approximation $\langle\psi_s |U_A(\vec{r},\vec{r}_0) |\psi_s\rangle\approx U_A(s,\vec{r}_0)$, where $s$ is the lateral separation with respect to the acceptor nucleus along the interface as shown in Fig.~\ref{prx_fig3}(b), yields
\begin{equation}\label{prx_eq:coulomb}
\Delta E_V \approx \frac{e_0^2}{4\pi\epsilon_{0}\epsilon_{\text{eff}}}\frac{1}{\sqrt{s^2+d^2}},
\end{equation}
where the modified dielectric constant $\epsilon_{\text{eff}}$ and depth $d$ can be independently determine from a fit \cite{Teichmann:2008bh,Lee:2010ko}. The modified dielectric constant at the interface $\epsilon_{\text{eff}} = (\epsilon_{\text{v}}+\epsilon_{\text{Si}})/2$ is due to the mismatch between the dielectric constants $\epsilon_{\text{v}}$ and $\epsilon_{\text{Si}} = 11.4$ \cite{Madelung:2004tb} of the vacuum and silicon, respectively, which leads to a single image charge at $-d$ as shown in Fig.~\ref{prx_fig3}(c). Figure~\ref{prx_fig3}(d) shows measured (filled squares) and fitted (solid line) spectral shift of the valence-band edge $\Delta E_V$ as function of lateral tip separation $s$ from the dopant. Following references \cite{Teichmann:2008bh,Lee:2010ko} we fit  the spectral shift of the valence-band edge as a function of position to equation~\ref{prx_eq:coulomb} (Fig.~\ref{prx_fig3}(d)) using the depth $d$ of individual acceptors and the modified dielectric constant $\epsilon_{\text{eff}}$ as two, independent, fitting parameters. Importantly, any screening by either by carriers in the tip or in the substrate would result in a shallower potential and thus lead to an increase of the modified dielectric constant from $\epsilon_{\text{eff}} = (\epsilon_{\text{v}}+\epsilon_{\text{Si}})/2$. The obtained modified dielectric constant for all five measured acceptors agree within experimental error with the expected value $\epsilon_{\text{eff}}=6$ following the classical half-space approach and experimental values that have previously been reported for STM experiments \cite{Teichmann:2008bh,Lee:2010ko}.\\



\begin{figure}
\includegraphics[width=90mm]{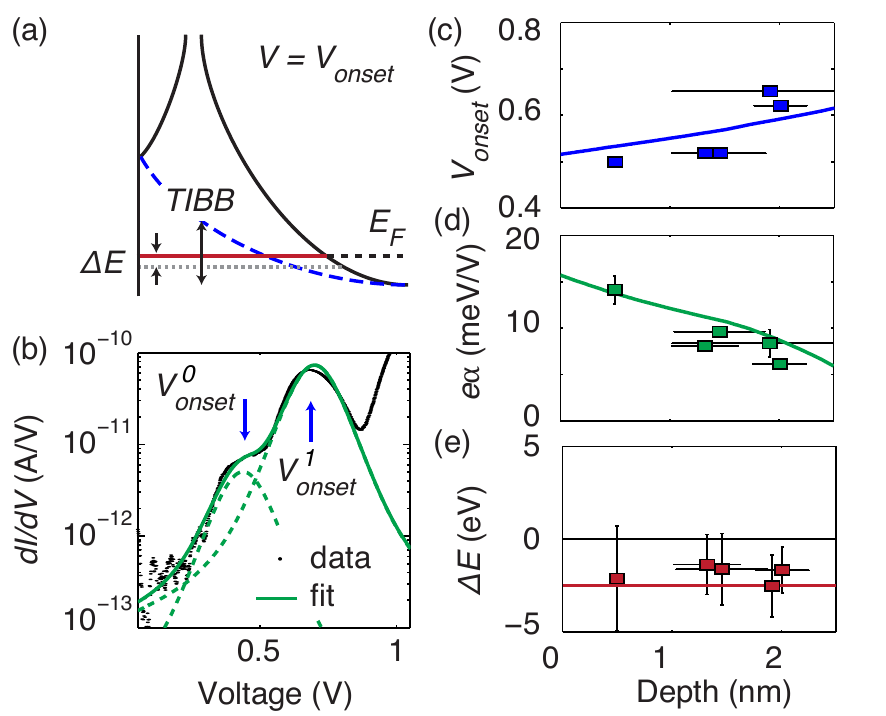}
\caption{\label{prx_fig4} (a) Schematic energy diagram. Local tip induced band bending brings the acceptor level $E_A$ into resonance with the Fermi energy. The voltage $V_{onset}$ at which this occurs depends on the acceptor depth, ionization energy and the screening length. (b) The lever arm $\alpha$ and the onset voltage $V_{onset}$ are determined fitting the two conductance peaks in the bandgap to the sum of two thermally broadened Lorentzian line shapes (relevant fitting parameters are listed in Appendix C).(c) Measured (squares) and calculated (line) onset voltage for resonant tunnelling as a function of acceptor depth. (d) Measured (squares) and calculated (line) lever arm as a function of acceptor depth.  (e) The shift in ionisation energy $\Delta E$ with respect to the bulk ionisation energy inferred from the measured onset voltage and the lever arm (squares). The onset voltage and lever arm are calculated for $\Delta E = -2.5$~meV, line in (c), and an overlap between the acceptor wavefunction and the tip induced band bending $\Delta E =  - 0.06\times$TIBB, where TIBB is the tip induced band bending at the acceptor site as illustrated in (a).}
\end{figure}

The lever arm $\alpha$, i.e. the shift of the acceptor energy levels $\Delta E$ due to the applied bias voltage, depends on the screening of the electric field in the semiconductor and the overlap between the acceptor wavefunction and the tip induced potential (Fig.~\ref{prx_fig4}(a)). Rather than trying to estimate the lever arm by solving the Poisson equation \cite{Feenstra:2003ua,Loth:2007us,Garleff:2010fp,Wijnheijmer:2011bg}, we fit the differential conductance peaks to a thermally broadened Lorentzian line shape\cite{Foxman:1993jq} (shown Fig.~\ref{prx_fig4}(b)): 
\begin{equation}
\label{eq:conv}
\begin{split}
dI/dV \propto &\int^{+\infty}_{-\infty}\cosh^{-2}(E/2k_BT) \\
 &\times \frac{\frac{1}{2}h\Gamma}{(\frac{1}{2}h\Gamma)^2+(\alpha e[V-V_{onset}] - E)^2}dE,
 \end{split}
\end{equation}
where $\Gamma=\Gamma_{in}+\Gamma_{out}$ is the sum of the tunnel-in and tunnel-out rates. This line shape describes resonant tunneling via a single, lifetime broadened, localized state into a continuum of states that are thermally occupied according to the Fermi-Dirac distribution \cite{Foxman:1993jq,Beenakker:1991uq,Averin:1991wo}. We observe two distinct differential conductance peaks within the band gap, due to the ground state and an excited state coming into resonance with the Fermi level of the substrate. Additional differential conductance peaks due to excited states are well understood within the framework of single-electron transport. Although we fit the both differential conductance peaks we will limit the discussion here to the energy level of the ground state, i.e. the binding energy of the acceptor. Charge noise can be excluded as dominant sources of conductance peak broadening as it does not have the appropriate line shape of the conductance peaks (see Appendix B). The lever arm $\alpha$ was allowed to vary linear in with $V$ in our fit by defining $\alpha=a+bV$ and making $a$ and $b$ independent fitting parameters. We observe that $bV$ is smaller than the confidence bounds on $a$ for all measured acceptors and thus that the measured values for $\alpha$ do not  depend on the bias voltage $V$. Consequently, we can conclude Stark shifts of the localized states due to the electric field is negligible in our measurement geometry. The squares in figure~\ref{prx_fig4}(c-e) indicate the measured voltage of ionization $V_{onset}$, lever arm $\alpha$ and extracted shift in binding energy with respect to the bulk bindingg energy $\Delta E$ as a function of depth for five different acceptors. The binding energy of all five acceptors are less than 5 meV smaller than the bulk binding energy.

For consistency we compare our method with the previously described method \cite{Feenstra:2003ua,Loth:2007us,Wijnheijmer:2009il,Garleff:2010fp,Wijnheijmer:2011bg} based on an electrostatic model. The method based on the electrostatic model consist of two steps: (i) first the voltage dependent potential near the interface is calculated using a 3D Poison model \cite{Feenstra:2003ua} (ii) then the shift of the dopant energy levels is estimated from the overlap between the dopant wavefunction and the potential near the interface (first-order correction using perturbation theory) \cite{Wijnheijmer:2009il}. The result can be expressed as $S\times \text{TIBB}$, where $S$ is the overlap integral and TIBB the potential at the dopant nucleus \cite{Wijnheijmer:2009il,Garleff:2010fp,Wijnheijmer:2011bg}. The two methods are compared by assuming $\Delta E=2.5$~meV (solid line in Fig.~\ref{prx_fig4}(e)) independent of depth and calculate back the corresponding $V_{onset}$ and $\alpha$ using the electrostatic model. The calculated values for $V_{onset}$ and $\alpha$ (solid lines in Fig.~\ref{prx_fig4}(c and d)) are then compared to the experimental values determined using our method (squares in Fig.~\ref{prx_fig4}(c and d)). We have used the overlap between the impurity wavefunction and the TIBB as a free parameter to match the calculated values for $V_{onset}$ and $\alpha$ (solid lines in Fig.~\ref{prx_fig4}(c-e)) to the experimental results (squares in Fig.~\ref{prx_fig4}(c-e)). We find that for $\Delta E=-0.06\times \text{TIBB}$ the Poisson model matches our results. Importantly, the 0.06 overlap is a factor $\sim$4 smaller the 0.27 overlap estimated in previous studies \cite{Garleff:2010fp,Wijnheijmer:2011bg}. This factor of 4 is consistent with the predicted uncertainty for the Poisson model \cite{Garleff:2010fp}.

Summarizing the main results, the squares in figure~\ref{prx_fig4}(c) indicate the onset voltage $V_{onset}$ of resonant tunnelling as a function of dopant depth for five different acceptors measured from the centre of the first differential conductance peaks. The measured lever arm $\alpha$ of the five acceptors, determined from the width of the differential conductance peaks, are indicated by the squares in Fig.~\ref{prx_fig4}(d). The shift in ionization energy $\Delta E = -\alpha e(V_{onset}-V_{FB})$ is smaller than zero for all measured acceptors, as indicated by the squares in Fig.~\ref{prx_fig4}(e). Since the measured value of $\epsilon_{\text{eff}}$ corresponds to the expected value following the classical half-space approach we can conclude that two opposing effects influence the ionization energy of near-interface dopant atoms: (i) dielectric mismatch; the dopant potential, which is screened by charge polarization in the semiconductor, becomes more attractive when its environment becomes less polarizable, i.e. has a lower dielectric constant, which leads to an increase of the ionization energy; (ii) quantum confinement; exclusion of the dopant wavefunction from the region outside the semiconductor results in a decrease of the ionization energy \cite{Hao:2009eq,Calderon:2010kz}. The observed bulk-like ionization energies for acceptors less than an effective Bohr radius from the interface are strong evidence that the effect of dielectric mismatch at the interface is mitigated by quantum confinement. Our transport data unambiguously demonstrates that acceptors within an effective Bohr radius from the interface of a silicon half-space geometry are not deactivated. We would like to point out that atomistic differences from the bulk may lead to an alteration of the binding energy such as the enhancement observed in the same geometry in Ref. \cite{Wijnheijmer:2009il}, but this effect is unrelated to dielectric mismatch.\\

\section{Conclusion}
In conclusion, we have determined the ionization energy of individual sub-surface acceptors below the Si(100):H surface by means of low-temperature scanning tunneling spectroscopy. Calibration of the local lever arm using the thermal broadening of the conductance peaks removes the necessity of modeling the tip induced electrostatic potential below the semiconductor surface, providing a direct measure for the ionization energy. We believe that this method provides a valuable parallel between extensively studied electron transport in mesoscopic devices and scanning tunneling spectroscopy. Moreover, this experiment demonstrates bulk-like ionization energies for acceptors less than a Bohr radius away from the interface. The fact that sub-surface acceptors are not deactivated is of great importance for the doping of nanoscale devices. 

\begin{acknowledgements}
This research was conducted by
the Australian Research Council Centre of Excellence for
Quantum Computation and Communication Technology
(project number CE110001027) and the US National Security
Agency and the US Army Research Office under contract
number W911NF-08-1-0527. M.Y.S. acknowledges an ARC
Federation Fellowship. S.R acknowledges an ARC Future Fellowship.
\end{acknowledgements}

\appendix
\section{Flat-band voltage and apparent barrier height}
In the main text the flat-band voltage is inferred from the position of the band edges after references \cite{Feenstra:1994ur,Feenstra:2003ua}. Here we verify the flat-band voltage from an independent measure of the apparent barrier height after references \cite{Loth:2007us,Wijnheijmer:2010uc}

\begin{figure}
\center
\includegraphics[width=90mm]{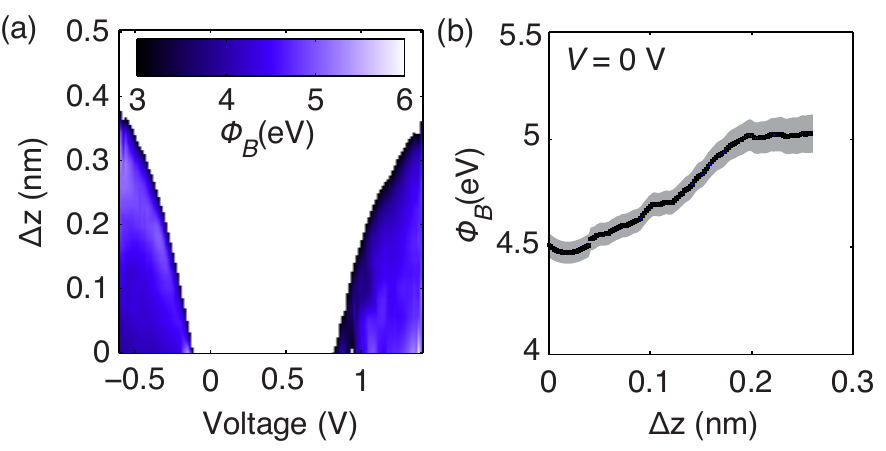}
\caption{\label{fig:bh}(a) Measured barrier height as a function of voltage and relative tip position $\Delta z$. The imaging condition $\Delta z=0$ corresponds to a tunnel current $I=300$~pA and bias voltage $V=-0.6$~V. (b) Extracted effective barrier height at $V=0$.}
\end{figure}

Figure~\ref{fig:bh}(a) shows the measured barrier height $\Phi_B$ as a function of bias voltage $V$ and relative tip-sample separation $\Delta z$. In approximation the tunnel current depends exponentially on the tip-sample separation $z$, $I \propto \exp(-2\kappa z)$ \cite{Feenstra:1987ic,Loth:2007us,Wijnheijmer:2010uc}, so that the inverse decay length can be determine by numerically differentiating the logarithmic current:
\begin{equation}
\kappa = -\frac{1}{2}\frac{\partial \ln I}{\partial z}.
\end{equation}
The apparent barrier height is determined from the inverse decay length using $\kappa = \sqrt{2m_0\Phi_B/\hbar}$. From geometrical arguments it follows that \cite{Feenstra:1987ic,Loth:2007us,Wijnheijmer:2010uc}
\begin{eqnarray}
eV<E_g: \Phi_B &=&  \frac{\Phi_{tip} + \chi + E_g + eV - |TIBB|}{2},\\
eV>E_g: \Phi_B &=&  \frac{\Phi_{tip} + \chi + E_g - eV + TIBB}{2},
\end{eqnarray}
where $\Phi_{tip}$ is the tip workfunction and $\chi$ and $E_g$ the electron affinity and band gap of the sample, respectively. We fit the measured barrier height for $eV<E_g$ to this relation to extract the apparent barrier height at zero bias, as is shown in Fig.~\ref{fig:bh}(b). At $eV > E_g$ the tunnel current contains a contribution of tunnelling from the tip into the conduction band as well as a contribution from tunnelling from the accumulation layer. As a consequence we can only extract the zero bias apparent barrier height from the tunnel current from the valence band, i.e. $eV < E_g$. 

The extracted apparent barrier height clearly decreases for smaller tip-sample separations as expected from the effect of image charges \cite{Simmons:1963jp,Feenstra:1987ic,Wijnheijmer:2010uc}. For $\Delta z > 0.2$~nm the apparent barrier height appears constant. Taking the value of this plateau $\Phi_B = 5.0\pm0.1$~eV to be the true trapezoidal barrier, i.e. assuming that image charge effects do no longer play a role for $\Delta z > 0.2$~nm and assuming the $TIBB$ goes to zero, we obtain a tip workfunction $\Phi_{tip}=4.8\pm0.1$~eV and a flatband voltage $V_{FB}=0.4\pm0.1$~V.
\section{Peak broadening and charge fluctuations}
The energy level $E$ of the localised impurity state, and therefore the conductance through the localised state,  depends on its electrostatic environment. The charging/discharging of a site near the localised impurity state will result in in a shift $\Delta E$ of the energy level and a change  in differential conductance $g(V) \rightarrow g(V+\Delta V)$, where $\Delta V=\Delta E/\alpha e$. Since not all charge is stationary, the conductance is expected to fluctuate in time, producing noise. When several fluctuators are acting simultaneously this will result in low-frequency ($1/f$) noise \cite{Dekker:1991ii,Liefrink:1999bo}. It is well known that for a pure $1/f$ noise signal the variance grows logarithmically with sampling time $t_s$ \cite{Brophy:1970hq,Keshner:1982fd,Hooge:2000gt}. Consequently, the conductance peaks are expected to broaden with increasing sampling time. 

If there is a cut-off time lower than 20 ms we would not observe a dependence of the peak width on integration time, but rather a dependence on tunneling current. We have found no significant dependence of the peak width on the tunneling current for different tip heights for 3 out of 5 acceptors (see $\alpha'$ in Table I of Appendix C) and can conclude that hopping of injected charge between impurities does not dominate to the width of the differential conductance peaks.

This is consistent with the fact that: (i) we see no discrepancy between the slope of the tail of the differential conductance peaks on a log-scale (slope $= 1/k_BT$) and the full width half max of the differential conductance peaks (FWHM $= 3.5k_BT$). Appreciable fast fluctuation would result in a FWHM $> 3.5k_BT$ inconsistent with the fit shown in Fig. 4b; (ii) the near-surface ($<$ 2-3 nm depth) dopants we observe are typically twenty or more nm apart, making it difficult for them to be brought into bias conditions of charge fluctuation.\\

\section{Fitting parameters}
The relevant fitting parameters and errors, including the depth and modified dielectric constant, are presented in Table I.
\begin{table*}
\centering
\begin{tabular}{|c|cc|ccccc|}
\hline
\# & $d$ (nm)& $\epsilon_{eff}$ & $V_{onset}^0$ (V)& $\alpha\times10^{3}$ & $\alpha'\times10^{3}$ & $\tfrac{1}{2}\hbar\Gamma/3.5k_BT$ & $R^2$\\
\hline
1 & $0.5\pm0.01$ & $6\pm0.1$ & $0.50\pm0.01$ &  $14.1\pm1.5$ & & $1.4\pm0.2$ & 0.99 \\
2 & $1.3\pm0.3$ & $6\pm1$ & $0.51\pm0.01$ &  $8.1\pm0.4$ & & $0.06\pm0.04$ & 0.96 \\
3 & $1.5\pm0.4$ & $8\pm2$ & $0.51\pm0.01$ &  $9.6\pm0.5$ &$8.9\pm0.2$& $0.52\pm0.08$ & 0.98 \\
4 & $1.9\pm0.9$ & $7\pm2$ & $0.65\pm0.01$ &  $8.4\pm1.0$ & $8.5\pm0.3$ & $0.55\pm0.26$ & 0.95 \\
5 & $2.0\pm0.2$ & $6\pm1$ & $0.62\pm0.01$ &  $6.1\pm0.3$ & $6.6\pm0.1$& $0.17\pm0.02$ & 0.98 \\
\hline
\end{tabular}
\caption{Fitting parameters for the five measured sub-surface acceptors presented in Fig. 4 of the main text. $\alpha'$ was measured on the same acceptors for a different tip-sample separation, resulting a factor $\sim2$ increase in resonant tunnelling current.}
\label{tab:myfirsttable}
\end{table*}
Horizontal error bars in Fig. 4, which is are uncertainties in depth $u_{d}$,  are directly obtained from the fitting the valence band edge. Vertical error bars in  Fig. 4, $u_{\Delta E} $, are obtained the standard method of propagation of errors and are dominated by the uncertainty in the flat band voltage.\\
%
\end{document}